 \pdfoutput=1
\documentclass[%
preprint,
 amsmath,amssymb,
 aps,
floatfix,
]{revtex4-1}
\pdfoutput=1
\usepackage{appendix}
\usepackage{amsmath}
\usepackage{graphicx}
\usepackage{epstopdf}
\usepackage{graphics}
\usepackage{epsfig}
\usepackage[byname]{smartref}
\usepackage{hyperref} 
\usepackage{txfonts}
\usepackage{morefloats}
\usepackage{tocloft}
\usepackage{graphicx}
\usepackage{dcolumn}
\usepackage{bm}
\usepackage{hyperref}

\begin{document}

\preprint{V1}

\title{The effect of edge and impurities sites properties on their localized states in semi-infinite zigzag edged 2D honeycomb graphene sheet}

\author{Maher Ahmed}
\affiliation{Department of Physics and Astronomy, University of Western Ontario, London ON N6A 3K7, Canada}
\affiliation{Physics Department, Faculty of Science, Ain Shams University, Abbsai, Cairo, Egypt}
\email{mahmed62@uwo.ca}
%



\begin{abstract}
In this work, the tridiagonal method \cite{Selim2011} is used to distinguish between edges
modes and area modes to study the edge sites properties effect on edge
localized states of semi-infinite zigzag 2D honeycomb graphene sheet. The results show a
realistic behavior for the dependance of edge localized states of zigzag
graphene on the edge sites properties which explaining the experimental
results of measured local density of states at the edge of graphene
\cite{Klusek2005}, while at the same time removing the inconsistence between
the semiconductor behavior found in the experimental data for fabricated GNRs
\cite{Wang3,Bing} and the expected theoretical semi-metallic behavior
calculated without considering the edge properties effect on the edge
localized states
\cite{PhysRevB.54.17954,PhysRevB.59.8271,JPSJ.65.1920,Neto1}.
\end{abstract}

\pacs{Valid PACS appear here}
\maketitle


\section{Introduction}\label{intro}
It is known that the 2D materials with zigzag edged
nanoribbons of 2D honeycomb lattice structure has peculiar flat localized
edge states at the Fermi level \cite{Selim2011,PhysRevB.54.17954,PhysRevB.59.8271,JPSJ.65.1920,Neto1}, which is a result of the zigzag geometry
effect on the particles hopping flow in its edges sites.

These edge states are known to be important due to their effect on the
electronic properties and as a consequence in a variety of future
applications of the famous Zigzag Graphene Nanoribbons (ZGNR)
\cite{Bing,Neto1}. The edge localized states are, in general, dependent upon
the ribbon size and purity of the sample \cite{C0NR00600A,Selim2011}. From a theoretical point of view, the edges states depends on the
probability for an electron to hop from a site in the edge to a bulk site, or
to impurity site in the neighborhood. The edge atoms have a different
coordination number from the bulk atoms, this leads to a different hopping
parameter between the edge atoms and the bulk one. Such a difference is not
usually considered in previous calculations for that ZGNR edge localized
states with different approaches
\cite{PhysRevB.54.17954,PhysRevB.59.8271,JPSJ.65.1920,Neto1}, these
calculations show inconsistencies with experimental results for all
fabricated GNRs that has semiconductor behavior \cite{Wang3,Bing} which is a
consequence of the absence of the flat edge states at Fermi level.

In \cite{Ahmed} we found that the tridiagonal method has an advantage in
studying the edge properties effects on the edge localized states due to its
ability to separate the edges modes from area modes in case of the 2D square
lattice.

Therefore in this work, the tridiagonal method is used to study the effect
of edge sites properties on the edge localized states of semi-infinite zigzag
2D honeycomb sheet as study case. The similarity between semi-infinite ZGNR
and semi-infinite antiferromagnetic as both two-sublattice structure guide us
to follow the tridiagonal method steps used in study the surface modes of
Heisenberg antiferromagnetic \cite{PhysRev.185.752,Costa2}. The method allow
us also to study the effect of impurities introduced substitutionally in
impurities localized states of the semi infinite ZGNR.


\section{Theory for edge states and impurity states}

\begin{figure}[t]
\centering
\includegraphics[scale =0.2]{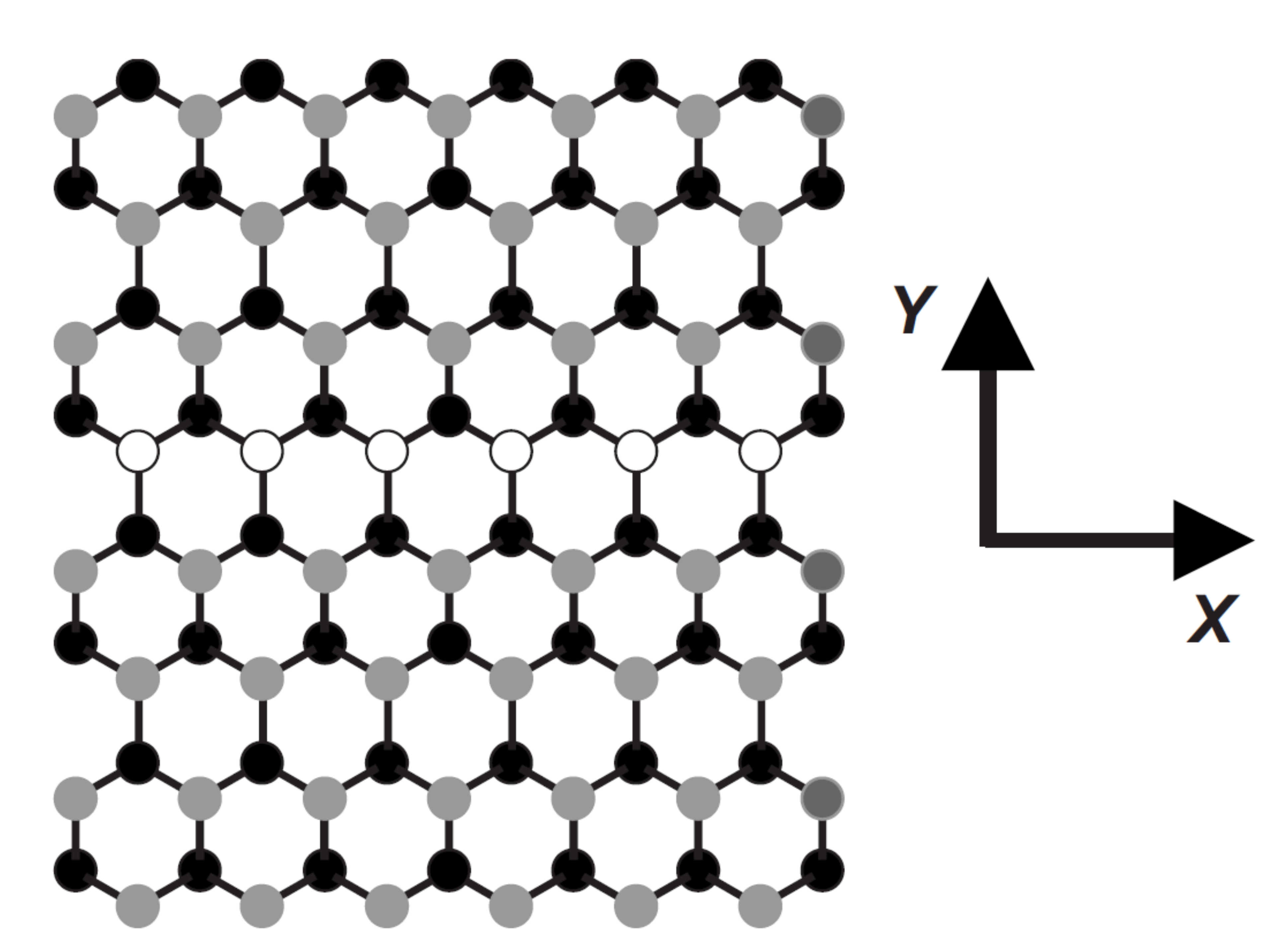}
\caption{Geometry of a graphene ribbon with zigzag edges. The black (gray) dots are the sublattice $A$ ($B$) atoms, where $A(B)$ sublattice type are labeled by index
$n(n')$ (= $1,2,\cdots,N$) and the
white dots show a row of impurities. Figure taken from \cite{rim1}.}
\label{Fig1}
\end{figure}

The structure of semi-infinite ZGNR is a honeycomb lattice of carbon atoms
with two sublattices denoted as $A$ and $B$. The geometry of a graphene
nanoribbon with zigzag edges is shown in Figure \ref{Fig1}, where the system
is infinite in the $x$ direction and has $2N$ rows of carbon atoms in the $y$
direction. To  be considered a ribbon, $N$ is a finite integer but here we
will extend the study for the the semi-infinite case where
$N\rightarrow\infty$. The $A(B)$ sublattice type lines are labeled with index
$n(n')$ where $n(n') = 1,2,3,\ldots$. The impurities (which may be silicon or
boron, for example) are introduced substitutionally along two different rows
of atoms parallel to the $x$ axis. The impurity lines, which preserve the
translational symmetry in the $x$ direction, may be any distance apart in the
ribbon.

\begin{table}[h]
 \caption{Nearest neighbor hopping matrix elements for the zigzag graphene nanoribbon}\label{tab:6}
\begin{tabular}{lc}
  \hline\hline
  Parameter& Zigzag \\ \hline
   $\beta$ &$2t \cos(\sqrt{3}q_x a/2)$\\
   $\gamma$& $t$  \\
  \hline
\end{tabular}
  \centering
\end{table}

Following a microscopic approach in terms of a tight binding model
Hamiltonian \cite{Neto1} with neglecting the next nearest
neighbor and with follow some recent work for impurities in graphene ribbons
\cite{Costa2} the Hamiltonian becomes
\begin{eqnarray}\label{ham1}
   H&=&-\sum_{i,j} t_{ij}  (a^\dag_{i}  b_{j} + a_{i}  b^\dag_{j})
\end{eqnarray}
where  $a^\dag_{i}$ (or $a_{i}$) creates (or annihilates) an electron on the
sublattice $A$ site, and $b^\dag_{j}$(or $b_{j}$) does the same for the
sublattice $B$ site, while $ t_{ij}$ is the nearest-neighbor hopping energy
between sublattices. In the pure material the hopping energy is denoted by
$t$ and its value is known \cite{Neto1} to be $\approx 2.8$ eV.

Taking into account the translational symmetry in $x$ direction, a Fourier
transform is made to rewrite Equation (\ref{ham1}) in a wavenumber
representation $q_x$ in the $x$ direction, and the rows are labeled $n$ and
$n'$. The Hamiltonian becomes

\begin{eqnarray}\label{ham2}
   H&=&\sum_{q_x,nn'}  \left[ \tau(q_x) a_{q_x,n}  b^\dag_{q_x,n'}+
   \tau(-q_x) a^\dag_{q_x,n}  b_{q_x,n'} \right].
\end{eqnarray}
The hopping amplitude factors $\tau_{nn'}(q_x)$ for the zigzag structure have
the form

\begin{eqnarray}\label{tau}
  \tau_{nn'}(q_x)=t\left[2\cos\left(\frac{\sqrt{3}}{2}q_xa\right)\delta_{n',n}
  +\delta_{n',n\mp1}\right],
\end{eqnarray}
or
\begin{equation}\label{tau1}
  \tau_{nn'}(q_x)=\beta\delta_{n',n}
  +\gamma\delta_{n',n\mp1},
\end{equation}
where the assignment of upper or lower signs depends on the sublattice type
sequence for rows $n$ and $n'$ (see Appendix \ref{AppB}). The definition of
$\beta$ and $\gamma$ is given Table \ref{tab:6}. Now we use the equation of
motion $i\hbar dX/dt=[X,H]$ for any operator $X$ for the creation and
annihilation operators of each row. Taking $\hbar=1$ and assuming that the
modes have a time dependence like $\exp[-i\omega(q_x)t]$, we obtain $2N$
coupled equations:

\begin{eqnarray}
\omega(q_x) a_{q_x,n}  &=&\sum_{q_x,n'}\tau_{nn'}(-q_x) b_{q_x,n'}   \nonumber\\
\omega(q_x) b_{q_x,n'}  &=&\sum_{q_x,n}\tau_{n'n}(q_x) a_{q_x,n'}. \label{equation}
\end{eqnarray}

Expanding and rearranging the Equations \eqref{equation} (see Appendix
\ref{AppB}) such that coupled equations between sublattice $A$ and sublattice
$B$ operators, could be written in the following form
\begin{eqnarray}\label{matrixw}
&-&   a_{q_x,n-1} +\frac{\left\{\omega^2(q_x) -\left(\beta^2 +\gamma^2\right)\right\}}{\beta\gamma} a_{q_x,n}- a_{q_x,n+1}=0 \nonumber \\
&& b_{q_x,n'} -\frac{\gamma} {\omega(q_x)} a_{q_x,n} -\frac{\beta}{\omega(q_x)} a_{q_x,n+1}=0
\end{eqnarray}
Equations \eqref{matrixw} could be written in the following supermatrix
equation \cite{PhysRev.185.752}

\begin{equation}\label{super1}
\left(
  \begin{array}{cc}
    A_N+\Delta A_N& O_N \\
    B_N  & I_N\\
  \end{array}
\right)
 \left(
  \begin{array}{c}
    a_N \\
    b_N \\
  \end{array}
\right)=0,
\end{equation}
where $O_N$ is the null matrix, $I_N$ the identity matrix, $a_N(b_N)$
operator column vector, and
\begin{equation}\label{super2}
A_N=\left(
  \begin{array}{cccccc}
    \zeta & -1   &     0 & 0 &0& \cdots \\
    -1   & \zeta & -1 & 0 & 0&\cdots \\
    0       & -1  &\zeta&  -1 &0  &\cdots \\
    0       &    0&  -1  & \zeta &  -1&\cdots \\
     \vdots & \vdots& \vdots & \vdots &\vdots &\ddots \\
  \end{array}
\right)
\end{equation}
and
\begin{equation}\label{bmat}
B_N=\left(
  \begin{array}{cccccc}
    \eta & 0   &     0 & 0 &0& \cdots \\
    \lambda   & \eta & 0 & 0 & 0&\cdots \\
    0       & \lambda &\eta&  0 &0  &\cdots \\
    0       &    0&  \lambda  & \eta &  0&\cdots \\
     \vdots & \vdots& \vdots & \vdots &\vdots &\ddots
  \end{array}
\right),
\end{equation}
the elements of above matrices are defined by
\begin{eqnarray}
  \zeta = \frac{\left\{\omega^2(q_x) -\left(\beta^2 +\gamma^2\right)\right\}}{\beta\gamma},\hspace{30pt}   \eta =  \frac{-\gamma} {\omega(q_x)}, \hspace{30pt}
 \lambda  =\frac{-\beta}{\omega(q_x)}.
 \end{eqnarray}

The edge properties have been separated from the area ``bulk" properties of
ZGNR by forming the matrix $\Delta A_N$. To simplify the calculations we
consider only putting one or two impurities lines in rows numbers $n_0$ and
$n'_0$ of sublattice $A$ such that their properties could be separated form
area properties in the same way the edge properties separated before by
including them in the matrix $\Delta A_N$. In this case the matrix $\Delta
A_N$ has the following form
\begin{eqnarray}
\Delta A_N= \left(
   \begin{array}{cccccccccccc}
 \Delta_e & \Delta_{s}  & 0 & 0 & 0 & 0 & 0 & 0 & 0 & 0 & 0 & \cdots\\
 \Delta_{s} & 0 & 0 & 0 & 0 & 0 & 0 & 0 & 0 & 0 & 0 & \cdots \\
     0 & 0 & 0 & 0 & 0 & 0 & 0 & 0 & 0 & 0 & 0 & \cdots \\
     0 & 0 & 0 &  0& \Delta_{In_0} & 0 & 0 & 0 & 0 & 0 & 0 & \cdots \\
     0 & 0 & 0 &  \Delta_{In_0} & \Delta_{n_0} & \Delta_{In_0}  & 0 & 0 & 0 & 0 & 0 & \cdots \\
     0 & 0 & 0 &  0 & \Delta_{In_0} & 0 & 0 & 0 & 0 & 0 & 0 & \cdots \\
     0 & 0 & 0 & 0 & 0 & 0 & 0 & 0 & 0 & 0 & 0 & \cdots \\
     0 & 0 & 0 & 0 & 0 & 0 & 0 & 0 & \Delta_{In'_0} & 0 & 0 & \cdots \\
     0 & 0 & 0 & 0 & 0 & 0 & 0 & \Delta_{In'_0} & \Delta_{n'_0} & \Delta_{In'_0} & 0 & \cdots \\
     0 & 0 & 0 & 0 & 0 & 0 & 0 & 0 & \Delta_{In'_0} & 0 & 0 & \cdots \\
     0 & 0 & 0 & 0 & 0 & 0 & 0 & 0 & 0 & 0 & 0 & \cdots \\
     \vdots & \vdots & \vdots & \vdots & \vdots & \vdots & \vdots & \vdots & \vdots & \vdots & \vdots & \ddots \\
   \end{array}
 \right), \label{matrixdfgh}
\end{eqnarray}
the elements of $\Delta A_N$ matrix are defined by
\begin{eqnarray}
  \nonumber
    \Delta_e &=& \zeta_e- \zeta,\hspace{30pt}    \Delta_{s} =\frac{\beta\gamma-\beta_e\tau_e}{\beta\gamma},\hspace{30pt} \zeta_e= \frac{\left\{\omega^2(q_x) -\left(\beta_e^2  +\tau_e^2\right)\right\}}{\beta\gamma}, \\
  \Delta_{n_0} &=&  \zeta_{n_0}-\zeta,\hspace{30pt} \Delta_{In_0}  =\frac{\beta\gamma-\beta_I\tau_I}{\beta\gamma},\hspace{30pt}
\zeta_{n_0}= \frac{\left\{\omega^2(q_x) -\left(\beta_{n_0}^2
+\tau_{n_0}^2\right)\right\}}{\beta\gamma}, \\
\Delta_{n'_0} &=&  \zeta_{n'_0}-\zeta,\hspace{30pt}\Delta_{In'_0}=\frac{\beta\gamma-\beta_{In'_0}\tau_{In'_0}}{\beta\gamma},\hspace{30pt}
\zeta_{n'_0}= \frac{\left\{\omega^2(q_x) -\left(\beta_{n'_0}^2  +\tau_{n'_0}^2\right)\right\}}{\beta\gamma}, \nonumber
\end{eqnarray}
where the edge hopping $t_e$, the first impurities line hoping $t_{n_0}$, and
the second impurities line hoping $t_{n'_0}$ have replaced the ZGNR interior
area sites hoping $t$ in the definition of $\beta$ and $\gamma$ in Table
\ref{tab:6} to obtain the edge and impurities counterpart.

Follow the steps of Heisenberg antiferromagnetic case \cite{PhysRev.185.752}
and the algebra of block matrices \cite{algebra}, one define the supermatrix
$G$

\begin{equation}\label{jdsjfhsdj}
G=\left(
  \begin{array}{cc}
    A_N& O_N \\
    B_N  & I_N\\
  \end{array}
\right)^{-1}= \left(
  \begin{array}{cc}
    (A_N)^{-1}& O_N \\
    -B_N (A_N)^{-1}  & I_N\\
  \end{array}
\right).
\end{equation}

Multiplying Equation \eqref{jdsjfhsdj} in  Equation\eqref{super1}, we get the
following
\begin{eqnarray}
 (I_N+(A_N)^{-1}\Delta A_N) a_N &=& 0 \nonumber\\
 (-B_n(A_N)^{-1}\Delta A_N) a_N +b_N&=& 0
\end{eqnarray}

Define the following matrix
\begin{equation}\label{Dm}
    D_N=I_N+(A_N)^{-1}\Delta A_N
\end{equation}
The matrix $D_N$ could be written in the following partition form (see
Appendix \ref{AppC})
\begin{equation}
    D_N= \left(%
\begin{array}{c|c}
 Q &  O \\ \hline
  S &    I \\
\end{array}%
\right),
\end{equation}
where $O$ is a square null matrix, $I$ a square identity matrix, $S$ a square
submatrix of $D_N$, and  $Q$ is square submatrix of $D_N$ with dimension of
$n'_0+1\times n'_0+1$.

The elements for the inverse of tridgional matrix $A_N$
\cite{CostaFilho2000195,Cottam1980,Cottam1976,PhysRev.185.752,PhysRev.185.720},
i.e. the matrix $(A_N)^{-1}$ is given as following:
\begin{equation}\label{inv:1}
((A_N)^{-1})_{nm}=\frac{x^{n+m}-x^{|n-m|}}{x+x^{-1}},
\end{equation}
where $x$ is a complex variable such that $|x|\leq 1$ and $x+x^{-1}=\zeta$.
As mentioned in \cite{Ahmed}, the values of $x$  should satisfy the following
boundary and physical conditions \cite{Cottam2004}. The area modes are
oscillating waves inside the nanoribbon, which requires that $x$ must be
imaginary exponential $x=e^{iq_y a/2}$ with $|x|=1$. From the definition of
$\zeta$ and $x$ parameters, the dispersion relation for the area band is
given by
\begin{eqnarray}
 \zeta&=&x+x^{-1} = e^{iq_y a/2}+e^{-iq_y a/2}=2 \cos (q_y a/2) \nonumber \\
    &=& \frac{\left\{\omega^2_B(q_x,q_y) -\left(\beta^2
+\gamma^2\right)\right\}}{\beta\gamma}\\
\omega_B(q_x,q_y) &=&\pm t\sqrt{1+4
\cos^2\left(\frac{\sqrt{3}q_x a}{2}\right)+ 4\cos\left (\frac{q_y a}{2}\right)\cos\left(\frac{\sqrt{3}q_x a}{2}\right)}\nonumber
\end{eqnarray}
This expression for the 2D area band for the zigzag nanoribbons is very
similar to the extended graphene electronic dispersion relation given in
\cite{Neto1}. This expression also
shows the same general features of graphene band structure \cite{Neto1}.

While the edge modes are localized on the edge and they are decaying
exponentially inside the nanoribbon, which requires that $x$ must be real and
less than 1 for edge modes. The edge modes are obtained by requiring the
determinant  of the coefficients for $a_n$ operator column vector to vanish
\cite{PhysRev.185.752,Cottam2004,RefWorks:21}:
\begin{equation}\label{det2}
    |D_N|=\det\left[\left(%
\begin{array}{c|c}
 Q &  O \\ \hline
  S &    I \\
\end{array}%
\right)\right],
\end{equation}
using the rules for obtaining the determinant of partitioned matrices
\cite{algebra}, Equation \eqref{det2} become
\begin{equation}\label{det3}
    |D_N|=|Q||I-SQ^{-1}O|=|Q|,
\end{equation}
the localized edge and impurities states, i.e. the edge and impurities
dispersion relations, for the semi-infinite zigzag are obtained by taking the
limit of Equation \eqref{det3} as $N\rightarrow\infty$
\begin{equation}\label{det4}
    \lim_{N\rightarrow\infty}|D_N|= \lim_{N\rightarrow\infty}|Q|=|Q|.
\end{equation}

In the case of taking only the edge properties effect on the edge localized
states, i.e there is no any impurities lines inside the sheet, Equation
\eqref{det4} become
\begin{equation}
   \det(Q)=\left|%
\begin{array}{cc}
 A^{-1}_{11} \Delta_{e}+ A^{-1}_{12}\Delta_{s}+1  & A^{-1}_{11} \Delta_{s} \\
 A^{-1}_{21} \Delta_{e}+ A^{-1}_{22}\Delta_{s}  &  A^{-1}_{21} \Delta_{s}+ 1\\
\end{array}%
\right|=0,\label{detref}
\end{equation}
which give
\begin{equation}\label{er}
\Delta_{s}^2x^{5}-2\Delta_{s}
x^{4}-(2\Delta_{s}^2+\Delta_{e})x^{3}
-(1-2\Delta_{s})x^2+(\Delta_{s}^2+\Delta_{e})x-1=0.
\end{equation}

In case the interaction of the edge with the interior sites is not affected
with the edge sites properties, i.e. $\Delta_{s}=0$, Equation \eqref{det4}
become
\begin{equation}\label{papser}
   \det(A)=\left|%
\begin{array}{cc}
 A^{-1}_{11} \Delta_e +1  & 0 \\
 A^{-1}_{21} \Delta_e  &   1\\
\end{array}%
\right|=A^{-1}_{11} \Delta_e+1=0
\end{equation}
which is the same expression obtained for Heisenberg antiferromagnetic
\cite{PhysRev.185.752}.
\begin{figure}[hp!]
\centering
\includegraphics[scale=1.2]{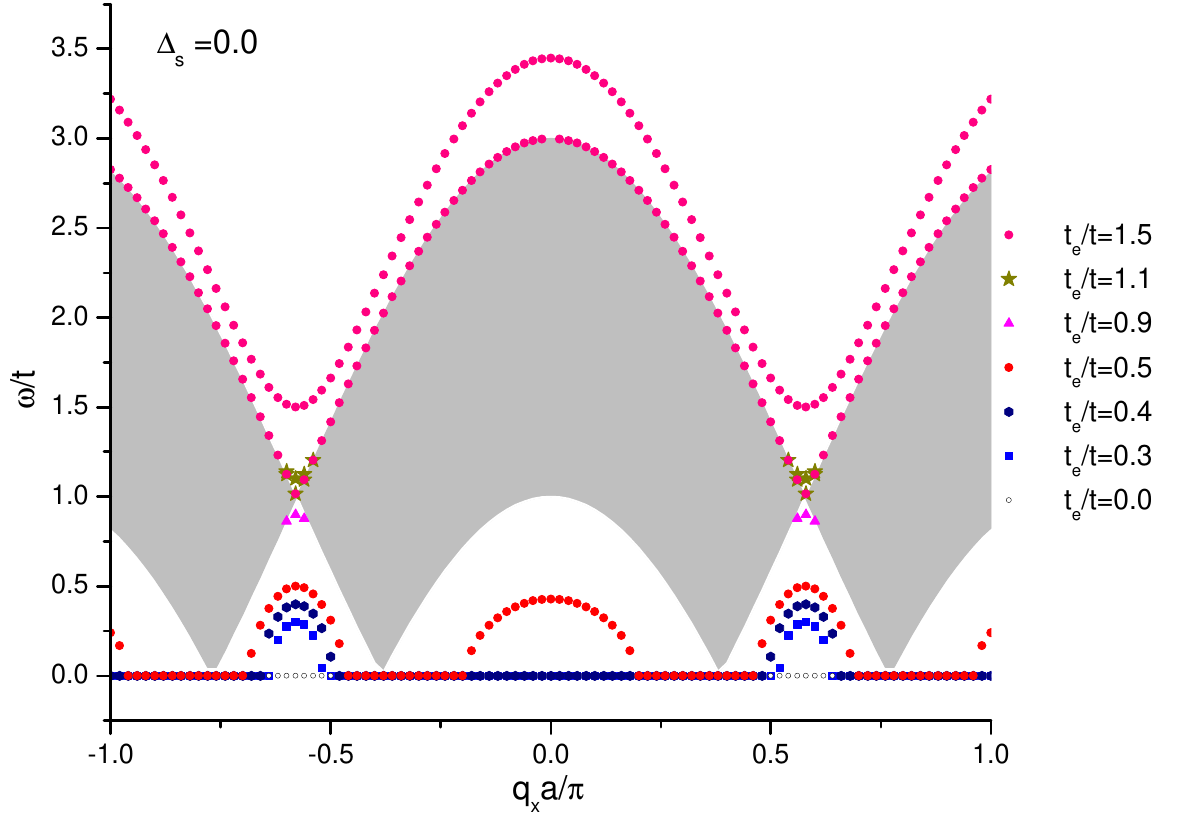}
\caption{Edge localized states for edge with different edge hopping calculated with $\Delta_s=0$, the shaded band represent area modes continuum.} \label{edges0}
\vspace{20pt}
\centering
\includegraphics[scale=1.2]{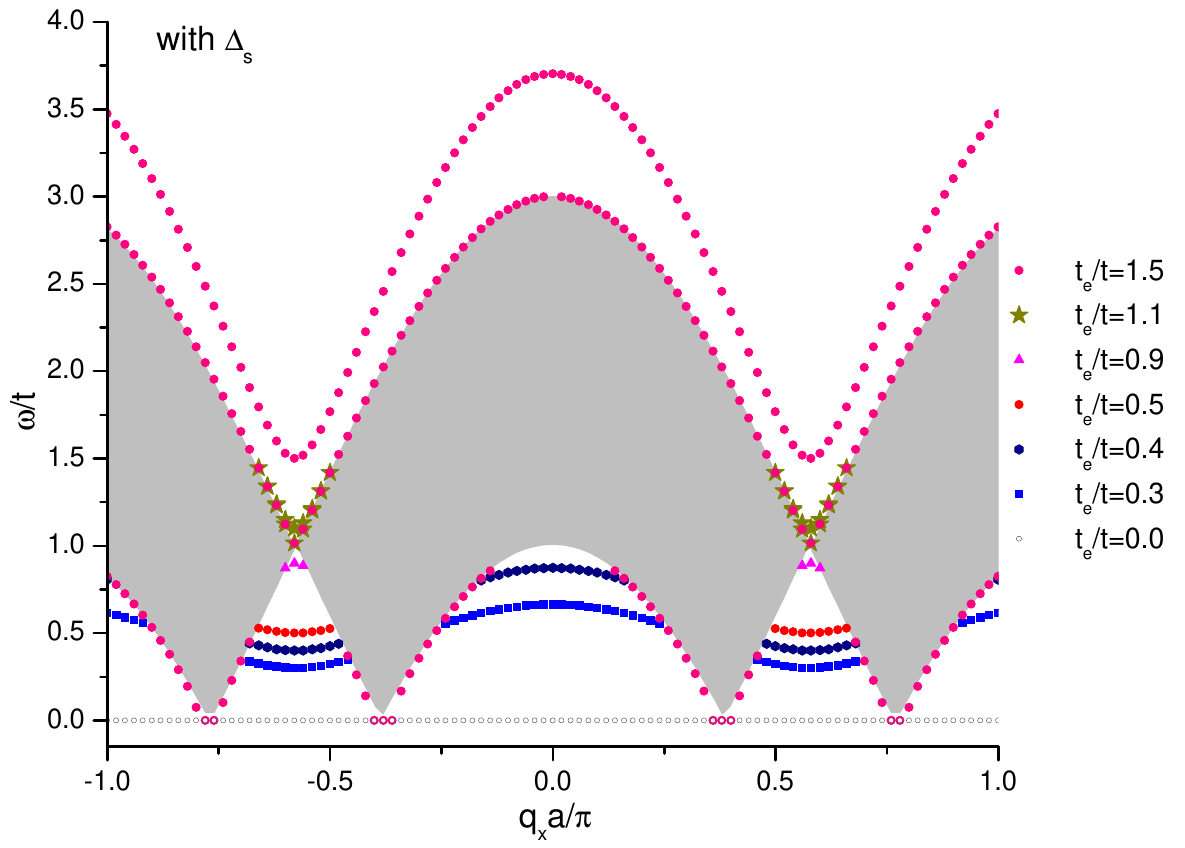}
\caption{Edge localized states for edge with different edge hopping calculated with $\Delta_s\neq0$, the shaded band represent area modes continuum.} \label{edgesn0}
\end{figure}

\section{Results}
Figures \ref{edges0} and \ref{edgesn0} show the edge localized states of
semi-infinite zigzag graphene sheet for different edge hopping to area
hopping ratios calculated with $\Delta_s=0$ using Equation \ref{papser} for
Figure \ref{edges0} and calculated with $\Delta_s\neq0$ using Equation
\ref{detref} for Figure \ref{edgesn0}. The Figures show that the dispersion
of the edge localized states depends on both the edge sites hopping
properties and their effect on the interaction with the interior sites in the
zigzag sheet.

The Figures begin with edge hopping equal to zero, which could be done by
saturating the carbons atoms on the edge. In this hopping value both
calculations (with and without $\Delta_s$) result in an extended flat
localized edge state through the whole Brilloin zone at Fermi level
$\omega_F/t =0$. This is due to the localized edge wave functions which agree
with density-functional theory (DFT) calculations for finite ribbon
\cite{xu:163102}.

As the edge hopping increases from zero, the dispersion of the edge localized
state begin to have a percentage of it laying at Fermi level $\omega_F =0$,
and the remaining percentage liftoff from the Fermi level. The percentage of
the edge dispersion that laying at Fermi level is important to the electronic
properties of the zigzag edged graphene nanoribbon.
\begin{figure}[h]
\centering
\begin{tabular}{cc}
\includegraphics[scale=.6]{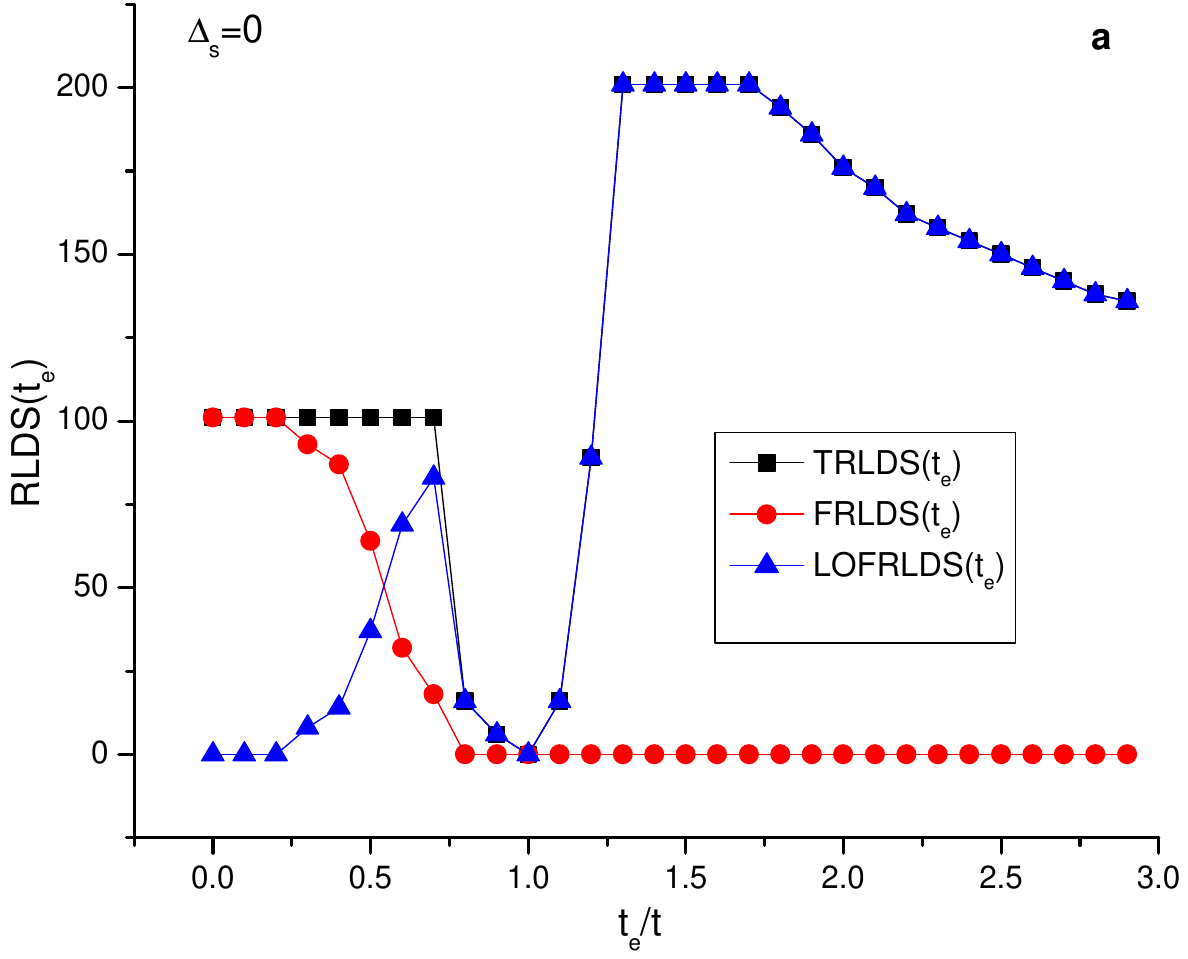}& \includegraphics[scale=.6]{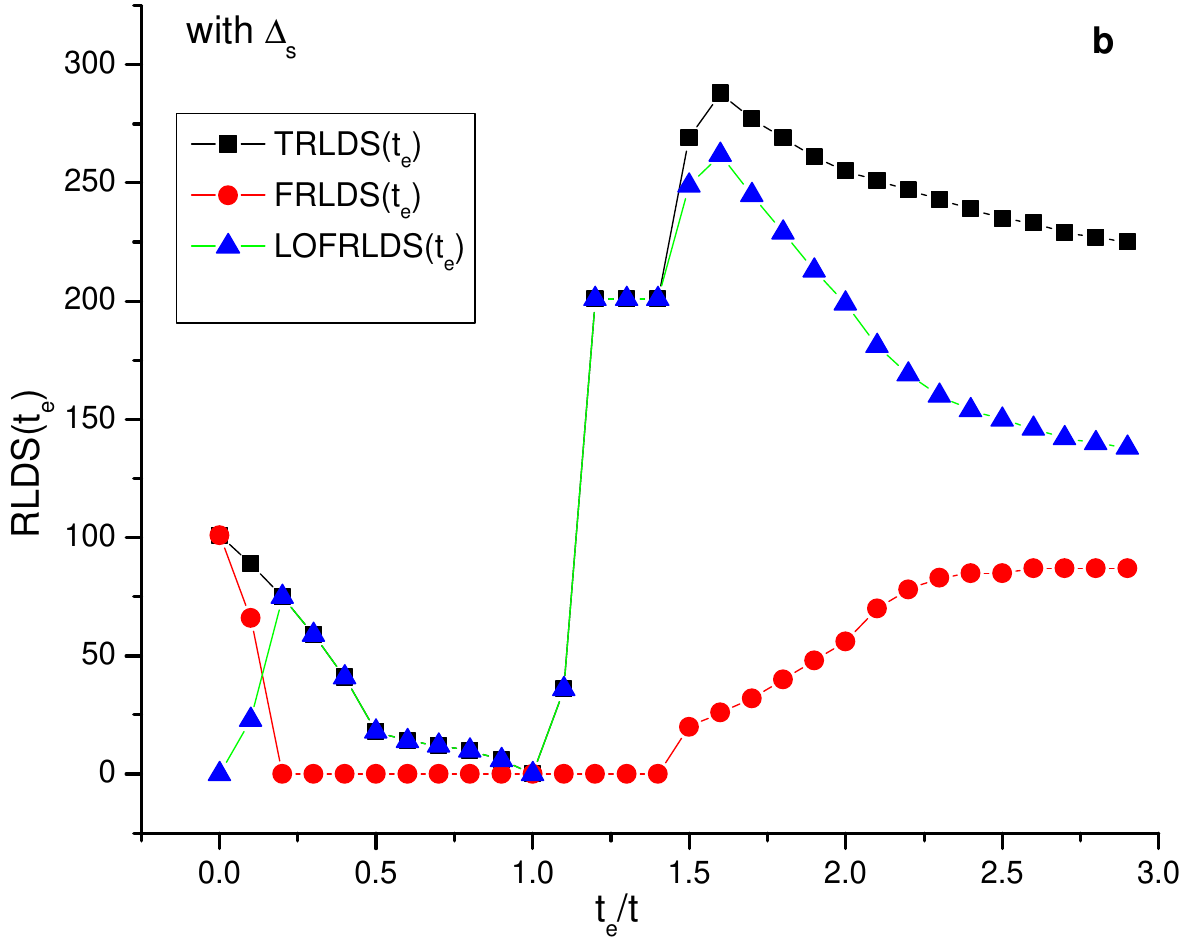}\\
\end{tabular}
\caption{The variation of TRLDOS, FRLDOS, and LOFRLDOS as a function
of edge hopping form 0 to 2.9. (a) for $\Delta_s=0$ and (b) for $\Delta_s\neq0$.}\label{TRLDOS5}
\end{figure}

To study the variation of the edge dispersions percentage laying at Fermi
level with the changing of the edge hopping and in the same time to easily
clarify the data displayed in Figures \ref{edges0} and \ref{edgesn0}, we
define the following three parameters for each given edge hopping value. The
first one is the Relative Localized Density of States near Fermi level
(FRLDOS), which is calculated computationally by counting the total number of
points in the localized edge dispersion with $\omega/t <0.2$. The second
parameter is the Total Relative Localized Density of States (TRLDOS), which
is calculated computationally by counting the total number of points in the
localized edge dispersion, which is a relative measure of its total density
of states. The third parameter is the difference between FRLDOS and TRLDOS
which represent the liftoff percentage of the edge localized state dispersion
from Fermi level (LOFRLDOS).

Figure \ref{TRLDOS5} shows the variation of TRLDOS, FRLDOS, and LOFRLDOS for
edge localized states as a function of edge to area hopping ratio form 0 to
2.9 with increment of 0.1, the (a) sub Figure represent the calculation with
$\Delta_s=0$, while (b) sub Figure represent the calculation with
$\Delta_s\neq0$.

With the help of the comparison between Figure \ref{TRLDOS5}(a) and Figure
\ref{edges0} and the comparison between Figure \ref{TRLDOS5}(b) and Figure
\ref{edgesn0}, the variation of the edge localized states dispersion with
different edge hopping values is described as follow:

First, for the dispersion of localized edge state calculated using $\Delta_s
=0$, starting with edge hopping equal to zero the Figure \ref{TRLDOS5}(a)
shows that FRLDOS is equal to TRLDOS which means that the edge localized
state lays completely in the Fermi level while LOFRLDOS is zero. In Figure
\ref{edges0} it is shown as a flat localized edge state at Fermi level
$\omega_F/t =0$ extended through the whole Brilloin zone as described above.
For 2 increment in the edge hopping there is no change in the value of FRLDOS
and it is still equal to TRLDOS with  LOFRLDOS is equal to zero, which mean
that dispersion did not change from as zero hopping. Beginning from edge
hopping equal to 0.3 the value of FRLDOS begin to decrease and LOFRLDOS begin
to increase while TRLDOS keep constant which means that some of dispersion
left off from Fermi level as shown in Figure \ref{edges0} as a left off near
$q_xa/\pi=\pm 0.5$. As edge hopping reach 0.5 the TRLDOS still keep constant
while FRLDOS still decreasing and LOFRLDOS still increases which means that
more dispersion left off from Fermi level as shown in Figure \ref{edges0} as
increase in the left off near $q_xa/\pi=\pm 0.5$ and around $q_xa/\pi=\pm
0.0$. At edge hopping 0.8 the TRLDOS drop quickly to small value and it is
equal to LOFRLDOS while FRLDOS become zero which means that most the edge
localized dispersion disappear which shown in edge hopping 0.9 in Figure
\ref{edges0} as small edge localized points near the intersection of the area
band segments. When edge hopping is equal to interior hopping, the edge
localized states completely disappear. Figure \ref{TRLDOS5}(a) shows that
FRLDOS is equal to zero in edge hopping range $0.9-2.9$. As the edge hopping
increases from 1 to 2.9, LOFRLDOS is equal to TRLDOS. The LOFRLDOS increases
very quickly to large value with increasing of edge hopping from 1 to 1.3,
and then keep constant in the edge hopping range $1.3-1.7$ after that range
LOFRLDOS decrease slowly. This behavior is shown in Figure \ref{edges0} as
increasing in the edge localized states above the area band.

In the second case, the dispersion of localized edge state calculated using
$\Delta_s\ne 0$, starting with edge hopping equal to zero the Figure
\ref{TRLDOS5}(b) shows that FRLDOS is equal to TRLDOS which means that the
edge localized state lays completely in the Fermi level while LOFRLDOS is
zero, in Figure \ref{edgesn0} it is shown as a flat localized edge state at
Fermi level $\omega_F/t =0$ extended through the whole Brilloin zone as
described above.  At the edge hopping equal to 0.1 the values of FRLDOS and
TRLDOS begin to decrease and LOFRLDOS begin to increase which means that some
of dispersion left off from Fermi level. In the edge hopping range from 0.2
to 1.4, the FRLDOS become zero and TRLDOS become equal to LOFRLDOS. In edge
hopping range 0.2 to 1, the LOFRLDOS decreases to a zero value, which is
shown in Figure \ref{edgesn0} as decreasing in the edge localized states
dispersion, and at the same time the edge localized state shifting up in the
energy. It is very important to note that at edge hopping 0.5, the edge
localized states dispersion become very similar to the famous peculiar edge
localized state for graphene zigzag nanoribbon
\cite{PhysRevB.54.17954,PhysRevB.59.8271,JPSJ.65.1920,Neto1} but here shifted
from Fermi level due to the edge hopping properties. When edge hopping is
equal to interior hopping, the edge localized states completely disappear.
The LOFRLDOS increases very quickly to large value with increasing of edge
hopping from 1 to 1.2, and then remains constant in the edge hopping range
$1.2-1.4$.  After that range, the LOFRLDOS increase to a peak at edge hopping
1.6 and then it begins to slowly decrease until edge hopping $2.9$. Starting
from the edge hopping 1.4, the LOFRLDOS begins to slowly increase with
increasing the edge hopping until the edge hopping reach  2.3, then the
LOFRLDOS begins converge to a nearly constant value. While the TRLDOS
increase to a peak at edge hopping 1.6 and it then begins to slowly decrease
with increasing the edge hopping. The behavior of the three parameters is
displayed in Figure \ref{edges0} as a change in the edge localized states
around the area band.
\begin{figure}[h]
\centering
\begin{tabular}{cc}
\includegraphics[scale=.6]{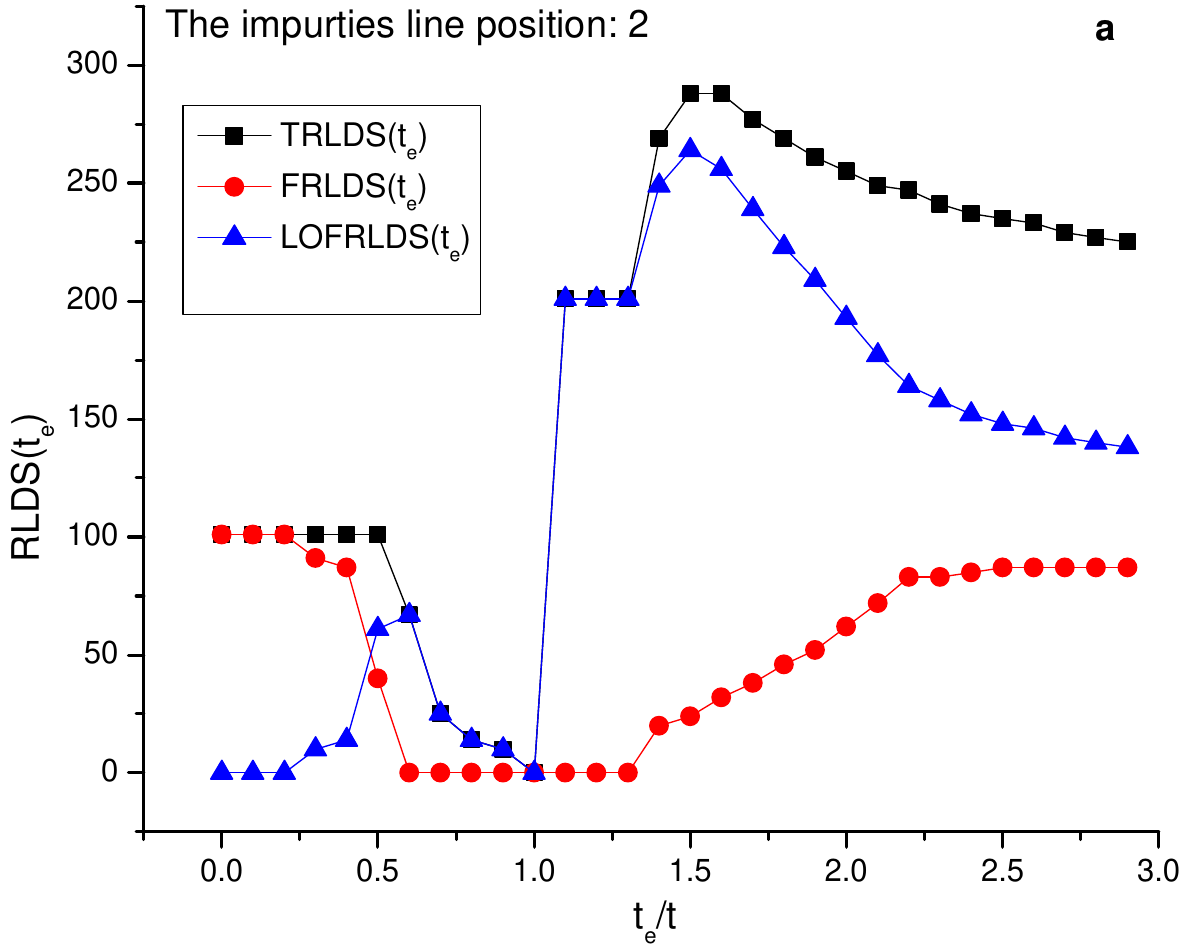}& \includegraphics[scale=.64]{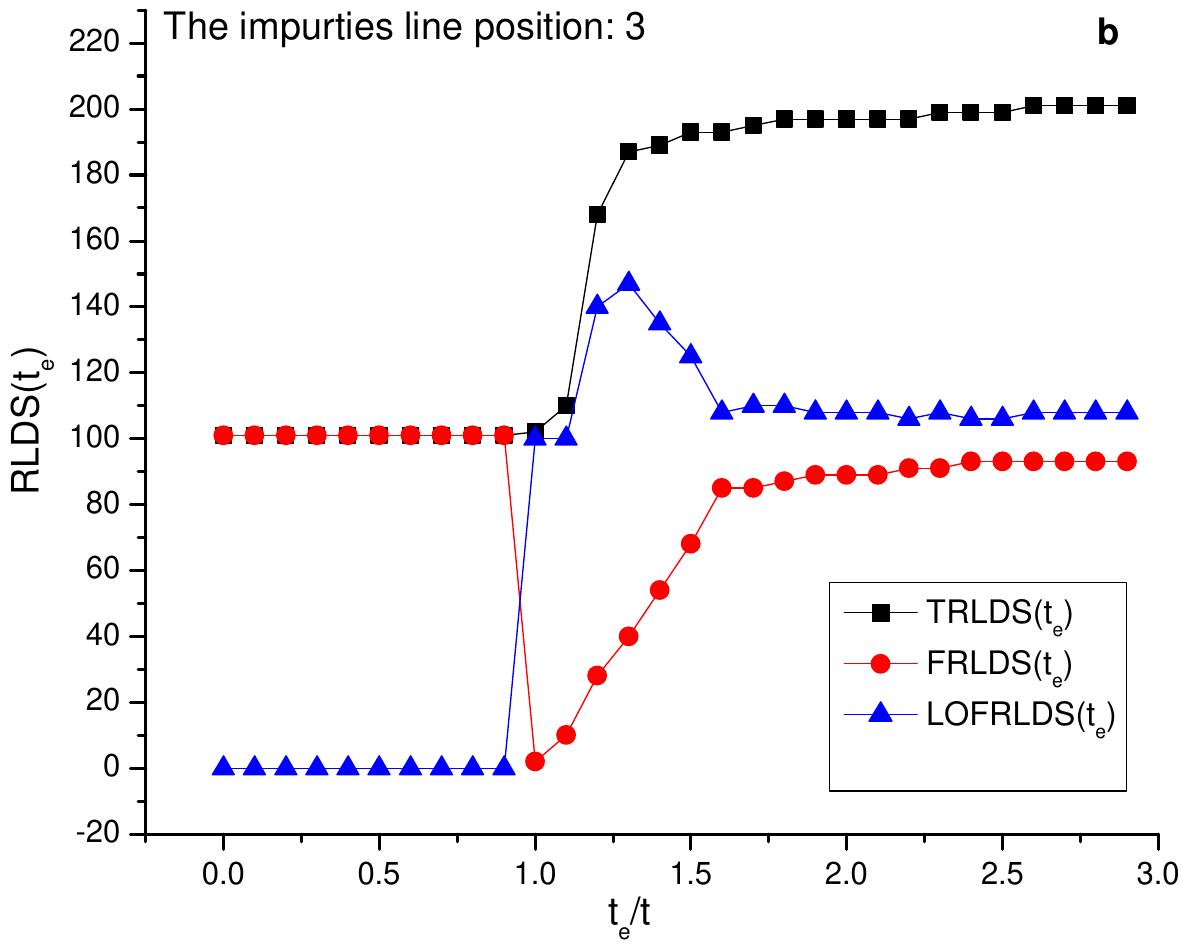}\\
\includegraphics[scale=.6]{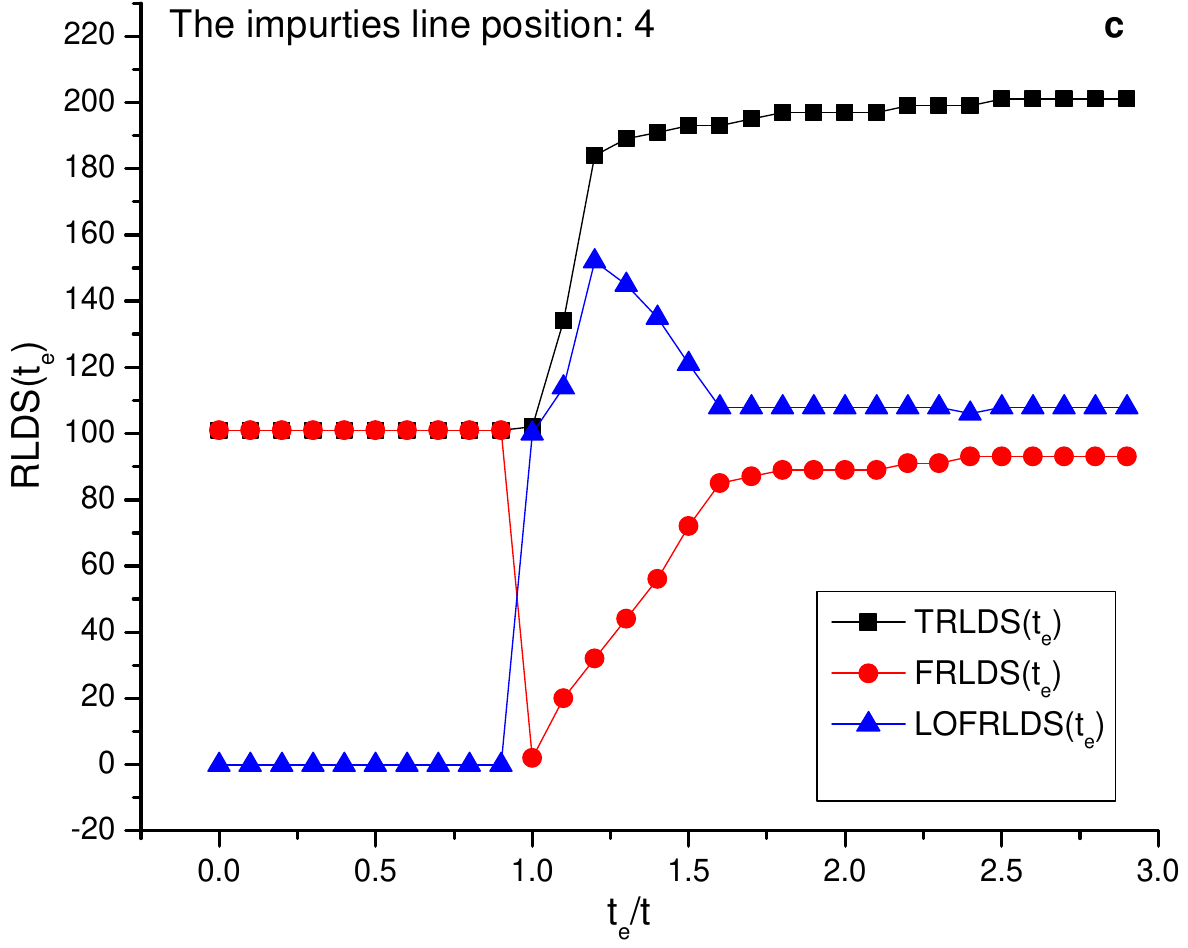}& \includegraphics[scale=.64]{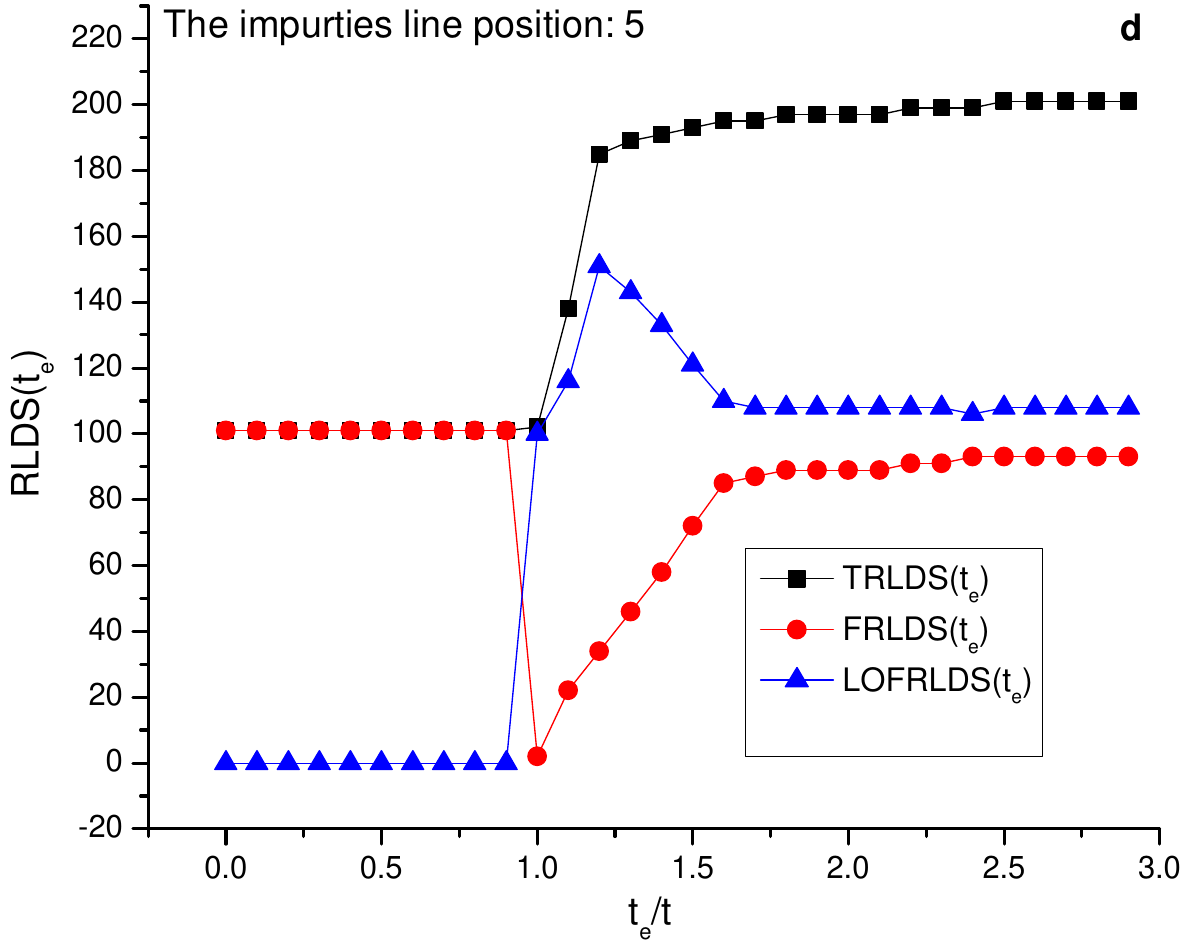}\\
\end{tabular}
\caption{The variation of TRLDOS, FRLDOS, and LOFRLDOS as a function of impurities hopping form 0 to 2.9 for the impurities line in sublattice $A$ at positions
(a) 2, (b) 3, (c) 4, (d)~5.}\label{TRLDOS5iln}
\end{figure}

The effects of both the impurities hopping and the impurities line position
on the impurities localized states have been calculated using Equation
\ref{det4} and the results are shown in Figures \ref{TRLDOS5iln}. This
calculation was done considering the edge hopping is equal to 1.

Figures \ref{TRLDOS5iln} show the variation of TRLDOS, FRLDOS, and LOFRLDOS
as a function of impurities hopping form 0 to 2.9 for the impurities line in
sublattice $A$ at positions (a) 2, (b) 3, (c) 4, (d)5. It is clear that the
dispersion of the impurities localized states for impurities line at second
raw of the sublattice $A$ is very similar to the dispersion of the edge
localized states described above. Beginning from position three in the
sublattice $A$, the dispersion of the impurities localized states becomes
completely different and nearly independent on the impurities line position
in the sublattice $A$.

Figures \ref{TRLDOS5iln}(a), (b), and (c) show that for the impurities
hopping range form 0 to 0.9 the FRLDOS is equal to TRLDOS which means that
the impurities localized state lays completely in the Fermi level while
LOFRLDOS is zero. In Figure \ref{edgesn0} it is shown as a flat localized
edge state at Fermi level $\omega_F/t =0$ extended through the whole Brilloin
zone as described above. At the impurities hopping equal to 1 the FRLDOS
switches to zero while LOFRLDOS changes to the value that keeps TRLDOS
constant. This means that the impurities localized state is completely lay
off from Fermi level with keeping its density of states constant. As
impurities hopping from 1 to 1.3 the LOFRLDOS rapidly increases to a peak in
its value, it then decreases until the impurities hopping reach 1.7 then the
LOFRLDOS begin converging to nearly a constant value. Likewise, as impurities
hopping from 1 to 1.7 the FRLDOS increase until the impurities hopping reach
1.7 the FRLDOS begin converging to nearly a constant value lower that
LOFRLDOS value. While the TRLDOS has fast increases in edge hopping range
from 1 to 1.3 and then it begin converging to a nearly high constant value.
\begin{figure}[h!]
\centering
\includegraphics[scale=1]{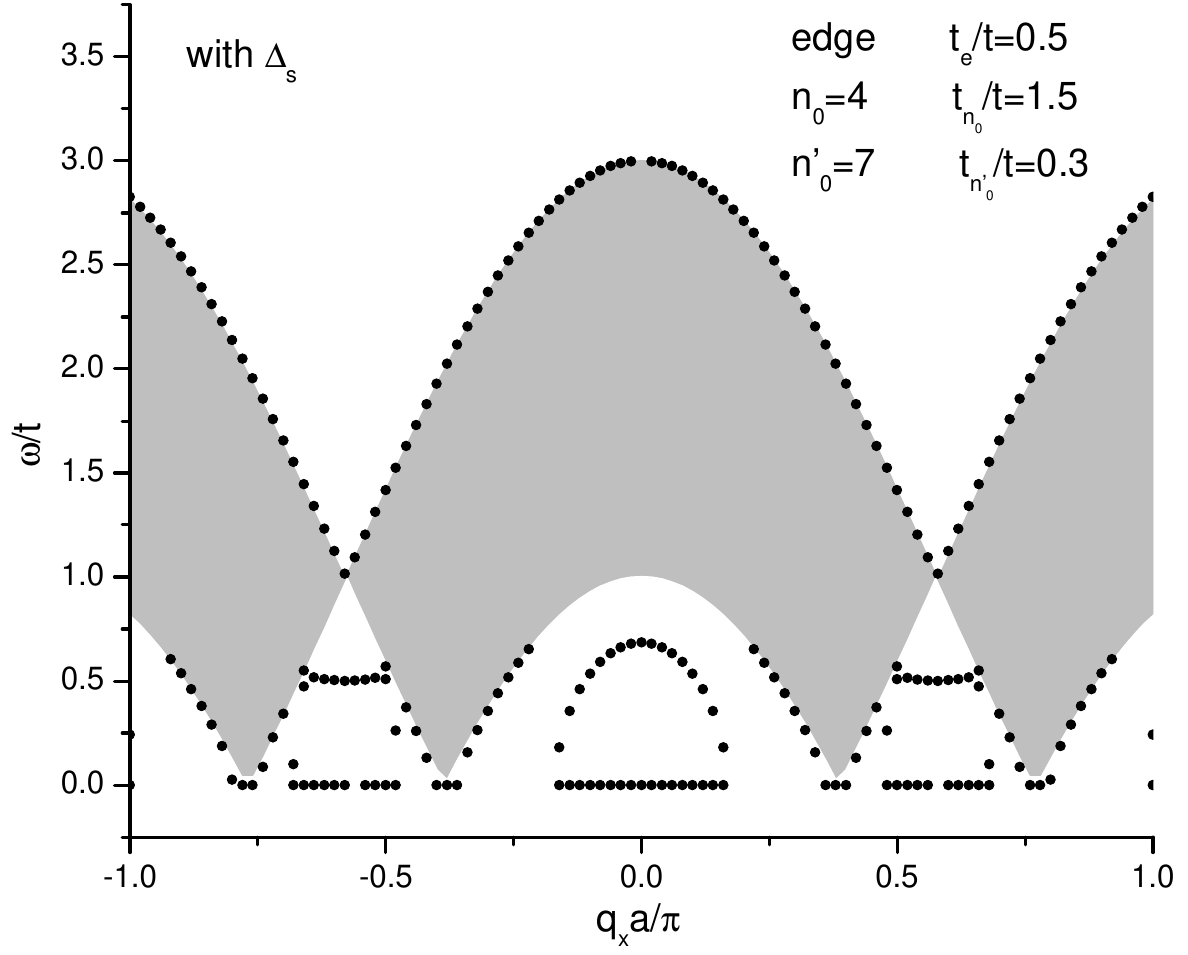}
\caption{Edge and impurities localized states the black dots for edge and two lines of impurities
at sublattice $A$, the shaded band represent area modes continuum. The Edge hopping is $t_e/t=0.5$,
the first impurities line position is  $n_0= 4$ with impurities hopping $t_{n_0}/t=1.5$, and
the second impurities line position is $n'_0= 7$ with impurities hopping $t_{n'_0}/t=0.3$.} \label{design6}
\end{figure}

\section{Discussion and Conclusions}
In this work the tridiagonal method was used to study the effect of edge
and impurities sites properties on their localized states in semi-infinite
zigzag edged 2D honeycomb sheet. It is found that the tridiagonal method
calculations provide us with two possibility to study the effect of the edges
sites properties on their localized states.  In the first one, the
interaction of the edge with the interior sites is not affected with the edge
sites properties, i.e. $\Delta_s=0$, and in the second one the interaction of
the edge with the interior sites is affected by the edge sites properties,
i.e. $\Delta_s\ne0$.

The results of the case $\Delta_s=0$ show that the edge localized states
dispersion has $q_x$ dependance of the hopping in 1D chain \cite{AlexanderAltland2010}
in low edge hopping values, which reflect the completely
isolation of edge hopping from interior one. This behavior is away from the
expected behavior from tight binding model results and the explanation given
in \cite{Selim2011}.

The results of the case $\Delta_s\ne0$ show that the edge localized states
dispersion has $q_x$ dependance of the hopping in 2D honeycomb but shifted in
the energy due to the edge hopping properties and especially at edge hopping
0.5, the edge localized states dispersion become very similar to the famous
peculiar edge localized state for graphene zigzag nanoribbon
\cite{PhysRevB.54.17954,PhysRevB.59.8271,JPSJ.65.1920,Neto1} but here shifted
from Fermi level. which reflect the importance of edge sites properties on
the edge hopping with interior sites. In this case the behavior is agree with
the expected behavior from tight binding model results and the explanation
given in \cite{Selim2011}.

In case of the effects of both the impurities hopping and the impurities line
position on the impurities localized states, the results show that second raw
of the sublattice $A$ has very similar edge localized states dispersion, and
after that position the impurities localized states become nearly independent
on the impurities line position in sublattice $A$. Also, the results show
that introducing impurities in any position of the sheet will produce
impurities localized states at Fermi level in all impurities hopping
properties, which affecting the electronic properties of the sheet.

The model could be used to study the effects of hopping properties
interaction for the edge and the two separated lines of impurities at
sublattice $A$ in the edge and the impurities localized states as shown in
Figure \ref{design6}.

The equivalent between the obtained mathematical expressions for edge states
in case of zigzag graphene and that obtained for surface spin wave in case of
Heisenberg antiferromagnetic \cite{PhysRev.185.752} reflect their equivalent
from geometrical and topological point of view. In the same time it show that
result is applicable to the magnetic and the 2D materials have the same
geometrical and topological structure.

Finally, the results of considering the interaction of the edge with the
interior sites is affected by the edge sites properties, i.e. $\Delta_s\ne0$,
show a realistic behavior for the dependance of edge localized states of
zigzag graphene on the edge sites properties which explaining the
experimental results of measured local density of states at the edge of
graphene \cite{Klusek2005}, and in the same time removing the inconsistence
between the semiconductor behavior found in the experimental data for
fabricated GNRs \cite{Wang3,Bing} and the expected theoretical semi-metallic
behavior calculated without considering the edge properties effect on the
edge localized states
\cite{PhysRevB.54.17954,PhysRevB.59.8271,JPSJ.65.1920,Neto1}.

\begin{acknowledgments}
This research has been supported by the Egyptian Ministry of Higher Education
and Scientific Research (MZA).
\end{acknowledgments}

\bibliography{xbib2}

\appendix

\section{Rearranging Equations \eqref{equation}}\label{AppB}
In this appendix we list the rearrange steps of Equations \eqref{equation} to obtain Equations \eqref{matrixw}.

\begin{eqnarray*}
\omega(q_x) a_{q_x,n}  &=&\sum_{q_x,n'}\tau_{nn'}(-q_x) b_{q_x,n'}   \nonumber\\
\omega(q_x) b_{q_x,n'}  &=&\sum_{q_x,n}\tau_{n'n}(q_x) a_{q_x,n'}.
\end{eqnarray*}
Expanding Equations \eqref{equation} using the sublattice indexes $n$ and $n'$
\begin{eqnarray*}
\omega(q_x) a_{q_x,1} &=&  \beta b_{q_x,1} \\
\omega(q_x) a_{q_x,2} &=& \gamma b_{q_x,1} +\beta b_{q_x,2} \\
\omega(q_x) a_{q_x,3} &=& \gamma b_{q_x,2} +\beta b_{q_x,3} \\
\vdots\\
\omega(q_x) a_{q_x,n} &=& \gamma b_{q_x,n'-1} +\beta b_{q_x,n'} \\
\end{eqnarray*}
\begin{eqnarray*}
\omega(q_x) b_{q_x,1} &=&\beta a_{q_x,1} +\gamma a_{q_x,2} \\
\omega(q_x) b_{q_x,2} &=& \beta a_{q_x,2} +\gamma a_{q_x,3} \\
\omega(q_x) b_{q_x,3} &=& \beta a_{q_x,3} +\gamma a_{q_x,4} \\
\vdots\\
\omega(q_x) b_{q_x,n'} &=& \beta a_{q_x,n} +\gamma a_{q_x,n+1} \\
\end{eqnarray*}
divide by $\omega(q_x)$, and rearrange we get
\begin{equation*}
     b_{q_x,n'} -\frac{\gamma} {\omega(q_x)} a_{q_x,n} -\frac{\beta}{\omega(q_x)} a_{q_x,n+1}=0
\end{equation*}

\begin{eqnarray*}
  \omega(q_x) b_{q_x,n'}   &=& \beta a_{q_x,n} +\gamma a_{q_x,n+1} \\
  \\
  \omega(q_x) b_{q_x,n'-1} &=& \beta a_{q_x,n-1} +\gamma a_{q_x,n} \\
\end{eqnarray*}
\begin{eqnarray*}
  \omega(q_x) a_{q_x,n} &=& \gamma b_{q_x,n'-1} +\beta b_{q_x,n'} \\
\\
\omega(q_x) a_{q_x,n}&=&\gamma \frac{\beta a_{q_x,n-1} +\gamma a_{q_x,n}}{ \omega(q_x)} +\beta \frac{\beta a_{q_x,n} +\gamma a_{q_x,n+1}}{ \omega(q_x)}\\
\\
\omega^2(q_x) a_{q_x,n}&=&\gamma (\beta a_{q_x,n-1} +\gamma a_{q_x,n} ) +\beta (\beta a_{q_x,n} +\gamma a_{q_x,n+1})
\end{eqnarray*}

lead to

\begin{equation*}
    -a_{q_x,n-1}+\frac{\omega^2(q_x)-(\beta^2+\gamma^2)}{\gamma \beta} a_{q_x,n}-a_{q_x,n+1}=0
\end{equation*}

\section{The partition of $D_N$ matrix}\label{AppC}

In this appendix we list the steps of partitioning the $D_N$ matrix.

\begin{eqnarray}
 A^{-1}= \left(
   \begin{array}{cccccccccccc}
 A^{-1}_{11} & A^{-1}_{12}  & A^{-1}_{13} & \cdots & A^{-1}_{1n_0-1} & A^{-1}_{1n_0} & A^{-1}_{1n_0+1} & \cdots & A^{-1}_{1n'_0-1} & A^{-1}_{1n'_0}& A^{-1}_{1n'_0+1}& \cdots\\
 A^{-1}_{21} & A^{-1}_{22}  & A^{-1}_{23} & \cdots & A^{-1}_{2n_0-1} & A^{-1}_{2n_0} & A^{-1}_{2n_0+1} & \cdots & A^{-1}_{2n'_0-1} & A^{-1}_{2n'_0}& A^{-1}_{2n'_0+1}& \cdots\\
A^{-1}_{31} & A^{-1}_{32}  & A^{-1}_{33} & \cdots & A^{-1}_{3n_0-1} & A^{-1}_{3n_0} & A^{-1}_{3n_0+1} & \cdots & A^{-1}_{3n'_0-1} & A^{-1}_{3n'_0}& A^{-1}_{3n'_0+1}& \cdots\\
\vdots & \vdots & \vdots & \vdots & \vdots & \vdots & \vdots & \vdots & \vdots & \vdots & \vdots & \cdots \\
 A^{-1}_{n1} & A^{-1}_{n2}  & A^{-1}_{n3} & \cdots & A^{-1}_{nn_0-1} & A^{-1}_{nn_0} & A^{-1}_{nn_0+1} & \cdots & A^{-1}_{nn'_0-1} & A^{-1}_{nn'_0}& A^{-1}_{nn'_0+1}& \cdots\\
     \vdots & \vdots & \vdots & \vdots & \vdots & \vdots & \vdots & \vdots & \vdots & \vdots & \vdots & \cdots \\A^{-1}_{n'_01} & A^{-1}_{n'_02}  & A^{-1}_{n'_03} & \cdots & A^{-1}_{n'_0n_0-1} & A^{-1}_{n'_0n_0} & A^{-1}_{n'_0n_0+1} & \cdots & A^{-1}_{n'_0n'_0-1} & A^{-1}_{n'_0n'_0}& A^{-1}_{n'_0n'_0+1}& \cdots\\
     \vdots & \vdots & \vdots & \vdots & \vdots & \vdots & \vdots & \vdots & \vdots & \vdots & \vdots & \ddots \\
   \end{array}
 \right)
\end{eqnarray}
which is of dimension $N\times N$.

\begin{eqnarray}
\Delta A_N= \left(
   \begin{array}{cccccccccccc}
 \Delta_e & \Delta_{s}  & 0 & 0 & 0 & 0 & 0 & 0 & 0 & 0 & 0 & \cdots\\
 \Delta_{s} & 0 & 0 & 0 & 0 & 0 & 0 & 0 & 0 & 0 & 0 & \cdots \\
     0 & 0 & 0 & 0 & 0 & 0 & 0 & 0 & 0 & 0 & 0 & \cdots \\
     0 & 0 & 0 &  0& \Delta_{In_0} & 0 & 0 & 0 & 0 & 0 & 0 & \cdots \\
     0 & 0 & 0 &  \Delta_{In_0} & \Delta_{n_0} & \Delta_{In_0}  & 0 & 0 & 0 & 0 & 0 & \cdots \\
     0 & 0 & 0 &  0 & \Delta_{In_0} & 0 & 0 & 0 & 0 & 0 & 0 & \cdots \\
     0 & 0 & 0 & 0 & 0 & 0 & 0 & 0 & 0 & 0 & 0 & \cdots \\
     0 & 0 & 0 & 0 & 0 & 0 & 0 & 0 & \Delta_{In'_0} & 0 & 0 & \cdots \\
     0 & 0 & 0 & 0 & 0 & 0 & 0 & \Delta_{In'_0} & \Delta_{n'_0} & \Delta_{In'_0} & 0 & \cdots \\
     0 & 0 & 0 & 0 & 0 & 0 & 0 & 0 & \Delta_{In'_0} & 0 & 0 & \cdots \\
     0 & 0 & 0 & 0 & 0 & 0 & 0 & 0 & 0 & 0 & 0 & \cdots \\
     \vdots & \vdots & \vdots & \vdots & \vdots & \vdots & \vdots & \vdots & \vdots & \vdots & \vdots & \ddots \\
   \end{array}
 \right),
\end{eqnarray}
which is of dimension $N\times N$.

\begin{equation}
    D_N=I_N+(A_N)^{-1}\Delta A_N
\end{equation}
therefore the $D_N$ dimension is $N\times N$ with the following elements:
\begin{eqnarray*}
D_{i1} &=&A^{-1}_{i1} \Delta_e+ A^{-1}_{i2}\Delta_{s}+\delta_{i1}\\
D_{i2} &=& A^{-1}_{i1} \Delta_{s}+ \delta_{i2}\\
D_{in_0-1}  &=&  A^{-1}_{in_0}\Delta_{In_0}+\delta_{in_0-1}\\
D_{in_0}    &=& A^{-1}_{in_0-1}\Delta_{In_0}+A^{-1}_{in_0}\Delta_{n_0}+A^{-1}_{in_0+1}\Delta_{In_0}+\delta_{in_0}\\
D_{in_0+1}    &=&  A^{-1}_{in_0}\Delta_{In_0}+\delta_{in_0+1}\\
D_{in'_0-1}  &=&  A^{-1}_{in'_0}\Delta_{In'_0}+\delta_{in'_0-1}\\
D_{in'_0}    &=& A^{-1}_{in'_0-1}\Delta_{In'_0}+A^{-1}_{in'_0}\Delta_{n'_0}+A^{-1}_{in'_0+1}\Delta_{In'_0}+\delta_{in'_0}\\
D_{in'_0+1}    &=&  A^{-1}_{in'_0}\Delta_{In'_0}+\delta_{in'_0+1}\\
D_{ij}&=& \delta_{ij}   \text{ if both $j$ and $i$ are not equal to either  $n$ or $m$}
\end{eqnarray*}

\begin{eqnarray*}
D_N = \left(
   \begin{array}{cccccccccccc}
 D_{11} & D_{12}  & D_{13} & \cdots & D_{1n_0-1} & D_{1n_0} & D_{1n_0+1} & \cdots & D_{1n'_0-1} & D_{1n'_0}& D_{1n'_0+1}& \cdots\\
 D_{21} & D_{22}  & D_{23} & \cdots & D_{2n_0-1} & D_{2n_0} & D_{2n_0+1} & \cdots & D_{2n'_0-1} & D_{2n'_0}& D_{2n'_0+1}& \cdots\\
D_{31} & D_{32}  & D_{33} & \cdots & D_{3n_0-1} & D_{3n_0} & D_{3n_0+1} & \cdots & D_{3n'_0-1} & D_{3n'_0}& D_{3n'_0+1}& \cdots\\
\vdots & \vdots & \vdots & \vdots & \vdots & \vdots & \vdots & \vdots & \vdots & \vdots & \vdots & \cdots \\
 D_{n_01} & D_{n_02}  & D_{n_03} & \cdots & D_{n_0n_0-1} & D_{n_0n_0} & D_{n_0n_0+1} & \cdots & D_{n_0n'_0-1} & D_{n_0n'_0}& D_{n_0n'_0+1}& \cdots\\
     \vdots & \vdots & \vdots & \vdots & \vdots & \vdots & \vdots & \vdots & \vdots & \vdots & \vdots & \cdots \\D_{n'_01} & D_{n'_02}  & D_{n'_03} & \cdots & D_{n'_0n_0-1} & D_{n'_0n_0} & D_{n'_0n_0+1} & \cdots & D_{n'_0n'_0-1} & D_{n'_0n'_0}& D_{n'_0n'_0+1}& \cdots\\
     \vdots & \vdots & \vdots & \vdots & \vdots & \vdots & \vdots & \vdots & \vdots & \vdots & \vdots & \ddots \\
   \end{array}
 \right)
\end{eqnarray*}

which give the following partition of $D_N$ matrix
\begin{equation}
    D_N= \left(%
\begin{array}{c|c}
 Q &  O \\ \hline
  S &    I \\
\end{array}%
\right),
\end{equation}
where $O$ is a square null matrix, $I$ a square identity matrix, $S$ a square
submatrix of $D_N$, and  $Q$ is square submatrix of $D_N$ with dimension of
$n'_0+1\times n'_0+1$.

\end{document}